\documentclass{ws-ijmpe}
\usepackage[super,compress]{cite}
\usepackage{multirow}
\usepackage{xcolor}
\usepackage{epsfig}
\usepackage{graphicx}
\usepackage{amsmath}
\usepackage{multirow}
\usepackage{tablefootnote}
\begin{document}

\markboth{H.~Garcilazo, A.~Valcarce, J.~Vijande}{Neutral baryonic systems with strangeness}

\catchline{}{}{}{}{}

\title{Neutral baryonic systems with strangeness}

\author{H. Garcilazo}

\address{Escuela Superior de F\' \i sica y Matem\'aticas, 
Instituto Polit\'ecnico Nacional, Edificio 9\\ 
07738 M\'exico D.F., Mexico\\
humberto@esfm.ipn.mx}

\author{A. Valcarce}

\address{Departamento de F\'\i sica Fundamental, Universidad de Salamanca\\ 
37008 Salamanca, Spain \\
valcarce@usal.es}

\author{J. Vijande}

\address{Departamento de F{\'\i}sica At\'omica, Molecular y Nuclear, 
Universidad de Valencia (UV) and IFIC (UV-CSIC),
46100 Valencia, Spain; \\
IRIMED Joint Research Unit (IIS La Fe - UV), 46100 Valencia, Spain\\
javier.vijande@uv.es}

\maketitle

\begin{history}
\received{Day Month Year}
\revised{Day Month Year}
\end{history}

\begin{abstract}
We review the status as regards the existence of three- and four-body bound 
states made of neutrons and $\Lambda$ hyperons. For interesting cases, the coupling 
to neutral baryonic systems made of charged particles of different strangeness has 
been addressed. There are strong arguments showing that the $\Lambda nn$ 
system has no bound states. $\Lambda\Lambda nn$ strong stable states are not 
favored by our current knowledge of the strangeness $-1$ and $-2$ baryon-baryon 
interactions. However, a possible $\Xi^- t$ quasibound state decaying to 
$\Lambda\Lambda nn$ might exist in nature. Similarly, there is a broad agreement
about the nonexistence of $\Lambda\Lambda n$ bound states. However, the coupling to
$\Xi NN$ states opens the door to a resonance above the $\Lambda\Lambda n$ threshold.

\end{abstract}


\ccode{PACS numbers:21.45.-v,25.10.+s,11.80.Jy}

\tableofcontents

\section{Introduction}

Bound states made of two neutrons or a neutron and a $\Lambda$ hyperon 
do not exist. Similarly, a $\Lambda\Lambda$ bound state, a core
element of the so-called $H$ particle~\cite{Jaf77}, has never been confirmed. 

The interest in bound states of neutrons and $\Lambda$ hyperons has been recently 
renewed by the experimental results of the HypHI Collaboration~\cite{Rap13}. 
They analyzed the reaction of $^6$Li projectiles at $2A$~GeV
on a fixed graphite target. Indications of a signal observed in the invariant 
mass distributions of $d+\pi^-$ and $t+\pi^-$ ($d$ stands for deuteron and $t$ for tritium)
final states were attributed to a strangeness-changing weak process corresponding to the 
two- and three-body decays of an unknown bound state of
two neutrons associated with a $\Lambda$, $^3_\Lambda n$, 
via $^3_\Lambda n\to t+\pi^-$
and $^3_\Lambda n\to t^\ast +\pi^- \to d + n+\pi^-$.

This is an intriguing conclusion since one would naively expect the $\Lambda nn$ system
to be unbound. In the $\Lambda nn$ system the two neutrons interact in the $^1S_0$
partial wave while in the $\Lambda np$ system, the hypertritium, they interact in 
the $^3S_1 - ^3D_1$ partial waves.
Thus, since the $NN$ interaction in the $J=0$ channel is weaker than in the $J=1$ channel,
and the $\Lambda np$ system is bound by only 0.13~MeV, one may have anticipated that the
$\Lambda nn$ system should be unbound. The unbound nature of the $\Lambda nn$ system was first 
demonstrated by Dalitz and Downs~\cite{Dal58,Dai58,Dal59} using a variational approach. 
In fact, the interpretation of the experimental data by the HypHI Collaboration
was immediately challenged by different theoretical
groups~\cite{Gar14,Hiy14,Gal14,Afn15}, supporting the longstanding 
outcome that the $^3_\Lambda n$ system is not bound~\cite{Dal58,Dai58,Dal59,Gar87}.

Bound states of two neutrons and two $\Lambda$ hyperons are another controversial subject.
Recently, Bleser {\it et al.}~\cite{Ble19} have offered a new interpretation 
of the results of the BNL AGS-E906 experiment to produce and study double hypernuclei through
a $(K^-,K^+)$ reaction on $^9$Be~\cite{Ahn01}.
Following a suggestion made by Avraham Gal, they explored the conjecture
that decays of a $^4_{\Lambda\Lambda}$n double hypernucleus may be 
responsible for some of the
observed structures in the correlated $\pi^- - \pi^-$ momenta.
A couple of theoretical groups~\cite{Ric15,Gar17} have discussed 
the possibility that a bound state may exist in the $\Lambda\Lambda nn$ system.
In particular, Ref.~\refcite{Gar17},
using local central Yukawa-type Malfliet-Tjon interactions,
concluded that the $\Lambda\Lambda nn$ system is unbound by 
a large margin. A recent calculation using the 
stochastic variational method in a pionless effective field theory 
approach~\cite{Con19} come to the identical conclusion.
It is important to notice that in order to create a $\Lambda\Lambda nn$ bound state the
four particles must coincide simultaneously since the system does not
contain two- or three-body subsystem bound states, so that the probability 
of the event occurring is rather small.

Thus, the possible existence of neutral baryonic systems
is a hot topic in nowadays strangeness nuclear physics.
In this brief review we will discuss first theoretical results as regards 
the $\Lambda nn$ system. Then, we will review those results related to 
the $\Lambda\Lambda nn$ system. We will present recent studies of the 
$\Lambda\Lambda nn - \Xi^- p nn$ coupled system.
Finally, we will address the $\Lambda\Lambda n$ system
and its coupling to $\Xi NN$ states before giving our conclusions.

\section{The $\Lambda nn$ ($^3_\Lambda n$) system}
\label{s2}

The unbound nature of the $\Lambda nn$ system was first
shown by Dalitz and Downs~\cite{Dal58,Dai58,Dal59} using a
variational approach. We review here the recent theoretical 
results demonstrating the non-existence of a bound 
$^3_\Lambda n$ state.

\subsection{Faddeev equations with separable potentials}
\label{ss21}
\begin{table}[b]
\tbl{Fredholm determinant at zero energy, $D_F(0)$, of the $(I,J^P)=(1,1/2^+)$ $\Lambda nn$ system for different 
models, A$-$D, of the $\Lambda n$ Nijmegen potential. They are characterized by the 
low-energy data, scattering length $a$ and effective range $r_0$ in fm, of the $^1S_0$ and $^3S_1$ channels.}
{\begin{tabular}{@{}p{1cm}cp{0.05cm}ccp{0.05cm}ccp{0.05cm}cp{1cm}@{}} 
\toprule
& \multirow{2}{*}{Model}  && \multicolumn{2}{c}{$^1S_0$} &&  \multicolumn{2}{c}{$^3S_1$} 
&& \multirow{2}{*}{$D_F(0)$} &   \\
& && $a_s$ & ${r_0}_s$ && $a_t$ & ${r_0}_t$ && & \\
\colrule
& A && $-$2.67 & 2.04 && $-$1.02 & 2.55 && 0.59 & \\
& B && $-$2.47 & 3.09 && $-$1.66 & 3.33 && 0.49 & \\
& C && $-$2.03 & 3.66 && $-$1.84 & 3.32 && 0.46 & \\
& D && $-$2.40 & 3.15 && $-$1.84 & 3.37 && 0.46 & \\ \botrule
\end{tabular}}
\end{table}

The non-existence of $\Lambda nn$ bound states 
was demonstrated by solving the three-body Faddeev equations with separable
potentials~\cite{Gar87}. The parameters of the two-body interactions 
were adjusted to reproduce the $\Lambda n$ scattering length and effective 
range obtained from four different versions of the Nijmegen 
potential~\cite{Nag73,Nag12,Nag77,Nag79}, see Table~1, as well as the 
nucleon-nucleon ($NN$) spin-singlet low-energy parameters
of Ref.~\refcite{Lac80}. In Refs.~\refcite{Gib74} and~\refcite{Gib80} it was shown
that when these potentials are replaced by separable interactions with the same 
low-energy parameters, they reproduce the hypertritium binding energy very accurately.
 
As pointed out in Ref.~\refcite{Gar87}, if a system can have at most 
one bound state then the simplest way to determine if it is bound or 
not is by looking at the Fredholm determinant, $D_F(E)$, at zero 
energy. If there are no interactions then $D_F(0)=1$, 
if the system in the overall is attractive then $D_F(0)<1$, and if a
bound state exists then $D_F(0)<0$. In the latter case, 
the energy of the bound state is obtained from the solution 
of the equation $D_F(E)=0$. In Ref.~\refcite{Gar87}
it was found that $D_F(0)$ for the $(I,J^P)=(1,1/2^+)$ $\Lambda nn$ system 
lies between 0.46 and 0.59 for the different models of the $\Lambda n$ interaction
constructed by the Nijmegen group, see Table~1, so that it 
is quite far from being bound.

\subsection{Faddeev equations with quark model-based interactions}
\label{ss22}

It could be argued that the use of simple separable potentials in Ref.~\refcite{Gar87} 
is not a realistic assumption. Besides, since 1987 the knowledge
of the strangeness $-1$ two-baryon interactions was improved and the models to 
study these systems were more tightly constrained. Thus, 
the $\Lambda nn$ system was reexamined in Ref.~\refcite{Gar14} with 
realistic baryon-baryon potentials obtained from the quark
model. 

The baryon-baryon interactions involved in the study of the coupled
$\Lambda NN - \Sigma NN$ system were obtained from a 
constituent quark cluster model (CQCM)~\cite{Val05,Vac05,Vij05}. In this model baryons
are described as clusters of three interacting massive (constituent) quarks,
the mass coming from the spontaneous breaking of chiral symmetry. The
first ingredient of the quark-quark interaction is a confining
potential. Perturbative aspects of QCD are taken into account
by means of a one-gluon exchange potential. Spontaneous breaking of 
chiral symmetry gives rise to boson exchanges
between quarks. 

In Refs.~\refcite{Gar07} and~\refcite{Gac07} the formalism
to study the $\Lambda NN$ system at threshold considering 
the effect of $D$ waves was established.
It leads to integral equations in two continuous variables,
the relative momentum of a pair and the relative momentum of
the third particle with respect to the pair. In order to solve 
these equations the two-body $t-$matrices are expanded in terms of Legendre 
polynomials leading to integral equations in only one continuous 
variable coupling the various Legendre components required for
convergence.
\begin{table}[t]
\tbl{Two-body $\Sigma N$ channels with a nucleon as spectator 
$(\ell_\Sigma \, s_\Sigma \, j_\Sigma \, i_\Sigma \, \lambda_\Sigma \, J_\Sigma)_N$,
two-body $\Lambda N$ channels with a nucleon as spectator 
$(\ell_\Lambda \, s_\Lambda \, j_\Lambda \, i_\Lambda \, \lambda_\Lambda \, J_\Lambda)_N$,
two-body $NN$ channels with a $\Sigma$ as spectator 
$(\ell_N \, s_N \, j_N \, i_N \, \lambda_N \, J_N)_\Sigma$, and
two-body $NN$ channels with a $\Lambda$ as spectator 
$(\ell_N \, s_N \, j_N \, i_N \, \lambda_N \, J_N)_\Lambda$ that contribute to
the $(I,J^P)=(1,1/2^+)$ $\Lambda NN - \Sigma NN$ state. $\ell$, $s$, $j$, and $i$, are, respectively, 
the orbital angular momentum, spin, total angular momentum, and isospin
of a pair, while $\lambda$ 
and $J$  are the orbital angular momentum of the third particle
with respect to the pair and the result of coupling $\lambda$ with the
spin of the third particle.}
{\begin{tabular}{@{}cccc@{}} 
 \toprule
$(\ell_\Sigma s_\Sigma j_\Sigma i_\Sigma\lambda_\Sigma J_\Sigma)_N$ &
$(\ell_\Lambda s_\Lambda j_\Lambda i_\Lambda\lambda_\Lambda J_\Lambda)_N$
& $(\ell_N s_N j_N i_N \lambda_N J_N)_\Sigma$
& $(\ell_N s_N j_N i_N \lambda_N J_N)_\Lambda$ \\
\colrule 
 (0\,0\,0\,1/2\,0\,1/2)  
& (0\,0\,0\,1/2\,0\,1/2)  
& (0\,0\,0\,1\,0\,1/2) & (0\,0\,0\,1\,0\,1/2) \\ 
  (0\,1\,1\,1/2\,0\,1/2)  
& (0\,1\,1\,1/2\,0\,1/2)  
& (0\,1\,1\,0\,0\,1/2) & \\ 
 (2\,1\,1\,1/2\,0\,1/2)  
& (2\,1\,1\,1/2\,0\,1/2)  
&  (2\,1\,1\,0\,0\,1/2) & \\
  (0\,1\,1\,1/2\,2\,3/2)  
& (0\,1\,1\,1/2\,2\,3/2)  
& (0\,1\,1\,0\,2\,3/2) & \\
 (2\,1\,1\,1/2\,2\,3/2) 
& (2\,1\,1\,1/2\,2\,3/2)
& (2\,1\,1\,0\,2\,3/2) &  \\ 
 (0\,0\,0\,3/2\,0\,1/2)  
& & &  \\ 
 (0\,1\,1\,3/2\,0\,1/2)  
& & & \\
 (2\,1\,1\,3/2\,0\,1/2)  
& & & \\
 (0\,1\,1\,3/2\,2\,3/2) 
& & & \\
 (2\,1\,1\,3/2\,2\,3/2)  
& & & \\
\botrule
\end{tabular}}
\end{table}
\begin{table}[b]
\tbl{Fredholm determinant at zero energy, $D_F(0)$,
of the $(I,J^P)=(1,1/2^+)$ $\Lambda NN - \Sigma NN$ state for
several hyperon-nucleon interactions characterized by
$\Lambda N$ spin-singlet, $a_{1/2,0}$, and spin-triplet, $a_{1/2,1}$,
scattering lengths (in fm).}
{\begin{tabular}{@{}ccccc@{}}
\toprule 
  & $a_{1/2,1}=-1.41$ & $a_{1/2,1}=-1.46$ & $a_{1/2,1}=-1.52$ 
& $a_{1/2,1}=-1.58$   \\
\hline
 $a_{1/2,0}=-2.33$ & 0.42 &  0.41 &  0.40 &  0.38 \\
 $a_{1/2,0}=-2.39$ & 0.42 &  0.41 &  0.39 &  0.38 \\
 $a_{1/2,0}=-2.48$ & 0.42 &  0.41 &  0.40 &  0.38 \\
\botrule
\end{tabular}}
\end{table}

The three-body problem was solved by 
taking full account of the $\Lambda NN - \Sigma NN$ coupling as 
well as the tensor force, responsible for the coupling between
$S$ and $D$ waves. In particular, for the $(I,J^P)=(1,1/2^+)$ $\Lambda NN$ channel 
which corresponds to the conjectured $\Lambda nn$
bound state, there are 21 coupled channels. 
To illustrate the completeness and complexity of the calculation, one shows
in Table~2 the quantum numbers of the contributing channels.

In Ref.~\refcite{Gar07} it was shown that increasing the $\Lambda N$ spin-triplet 
scattering length the $(I,J^P)=(0,3/2^+)$ $\Lambda NN$ state becomes
bound. Given that this state does not exist in nature, a lower limit of $-$1.58~fm 
was set for the $\Lambda N$ spin-triplet
scattering length. Since, in addition, the fit of the 
hyperon-nucleon cross sections is worsened
when the spin-triplet scattering length is larger than 
$-$1.41~fm~\footnote{Note that the signs of the $\Lambda N$ scattering lengths 
have been changed with respect to the original reference~\cite{Gac07}
in the text and in Tables~3 and~14, to have the same convention throughout the review.}
it was concluded that
$-1.41 \ge  a_{1/2,1} \ge -1.58$~fm. By requiring that the 
hypertritium binding energy has the experimental value 
$B=0.13\pm 0.05$~MeV, the following $\Lambda N$ spin-singlet scattering length
limits were obtained: $-2.37 \ge a_{1/2,0} \ge -2.48$~fm.
Thus, twelve different models corresponding to different choices of the spin-singlet 
and spin-triplet $\Lambda N$ scattering lengths were constructed. All of them describe 
equally well the available experimental data. We show in Table~3 the Fredholm determinant at zero energy of the 
$(I,J^P)=(1,1/2^+)$ $\Lambda NN - \Sigma NN$ state for these models. The realistic quark model interactions predict
a Fredholm determinant at zero energy ranging between 0.38 and 0.42, close to the
interval $0.46 - 0.59$ obtained from the separable potentials of the Nijmegen group
in Sect.~\ref{ss21}, see Table~1.
As one can see, in all cases the Fredholm determinant
at zero energy is positive and far from zero, excluding the possibility
of binding for this system. From the results of
Table~3 and from the energy dependence of the Fredholm
determinant shown in Fig.~2 of Ref.~\refcite{Gar07} one can infer that
the $(I,J^P)=(1,1/2^+)$ state is unbound by at least $5 - 10$~MeV, which is a large
energy in comparison with the 0.13~MeV binding energy of the hypertritium.

Thus, using either simple separable potentials or
a full-fledged calculation with realistic 
baryon-baryon interactions derived from a constituent quark cluster
model, there does not seem to be any possibility of 
existence of a $\Lambda nn$ bound state. 

\subsection{Constraints from $^3_\Lambda$H, $^4_\Lambda$He and $^3$H}
\label{ss23}

In Ref.~\refcite{Hiy14} the existence of the $^3_\Lambda n$ system
was scrutinized using a hyperon-nucleon ($YN$) potential
equivalent to the Nijmegen NSC97f interaction~\cite{Rij99}.
The model contains central, spin-orbit and tensor 
terms. For the $NN$ interaction the AV8 potential discussed in
Ref.~\refcite{Pud97}
was used. This interaction was modified
to describe a $^3_\Lambda n$ bound state, in order to see
what this modification does to other states that 
are well described by the original model.

The $^3_\Lambda n$ $J^P=1/2^+$ ground state includes $YN$
spin-singlet and spin-triplet contributions, while the $^3_\Lambda$H 
$J^P=1/2^+$ ground state is dominated by the $YN$ spin-singlet interaction.
Therefore, the spin-triplet interaction of the $YN$ interaction was tuned in
a manner that does not affect the binding energy of $^3_\Lambda$H
significantly. For this purpose the strength of the tensor
part of the $\Lambda N-\Sigma N$ coupling was multiplied by a fudge factor 1.2, 
because the tensor part of the $\Lambda N - \Sigma N$ coupling acts only in the 
spin-triplet $YN$ interaction. By doing this, however,
the $^3_\Lambda$H is overbound with a binding energy of 0.72~MeV to be 
compared with the experimental data, 0.13~MeV. In addition, an excited $J^P= 3/2^+$
state of $^3_\Lambda$H appears, with a binding energy of 0.43~MeV, 
for which there is no experimental evidence.

Moreover, the $^4_\Lambda$He $J^P=0^+$ ground state, which has an experimental binding energy of
2.21\,MeV and had a binding of 2.31~MeV in the original model, 
becomes bound by 5.55~MeV. On the other hand,
the $^4_\Lambda$He $J^P=1^+$ excited state, which has an experimental binding energy of
1.08~MeV and had a binding of 0.57~MeV in the original model, becomes
bound by 4.29~MeV.
 
Finally, the effect of varying the strength of the $NN$
$^1S_0$ interaction was also studied. If this interaction 
is multiplied by a factor 1.35, the $^3_\Lambda n$ system gets bound 
by 1.272~MeV. However, the $nn$ subsystem also becomes bound 
by 1.269~MeV and the $^3$H, which had 
a binding energy in the original model of 7.77~MeV, 
becomes overbound with a binding energy of 13.93~MeV.

\subsection{Constraints from $\Lambda p$ scattering, $^3_\Lambda$H,
and $^4_\Lambda$H}
\label{ss24}

In Ref.~\refcite{Gal14} the nonexistence of the $^3_\Lambda n$ system
was demonstrated by using Yamaguchi separable potential models~\cite{Yam54}
of the $NN$ and $YN$ interactions. For a Yamaguchi separable potential the range $\alpha$
and strength $\gamma$ are completely determined by the 
low-energy parameters, the scattering length $a$ and the effective range $r_0$. 
Thus, neglecting the spin dependence of the $YN$ interaction and using reasonable
values for the effective range such as $r_0=2.5$~fm or $r_0=3.5$~fm, see Table~1,
the $YN$ scattering lengths required to give a $^3_\Lambda$H
binding energy of $0.13$~MeV are, respectively, $a=-1.498$~fm and
$a=-1.895$~fm. 

These low energy parameters give rise to a $\Lambda p$ cross 
section at $p_\Lambda=145$~MeV/c of $\sigma_{\Lambda p}=$ 192.5~mb 
and 239.7~mb, respectively, close to the experimental value of 180~mb~\cite{Ale68}. 
On the other hand, assuming a $^3_\Lambda n$ state with zero binding 
energy, it leads to scattering lengths $a=-4.492$~fm and $a=-5.930$~fm, respectively, which in 
turn give $\sigma_{\Lambda p}=953.8$~mb and 
$\sigma_{\Lambda p}=943.1$~mb, respectively, in strong disagreement
with the experiment. These values of $a$ lead also to $^3_\Lambda$H 
binding energies of 2.59 and 1.74~MeV, respectively, in complete disagreement
with the experimental value of 0.13~MeV.
\begin{table}[t]
\tbl{Binding energy $B(^2n)$ (in MeV) of two neutrons in a separable
Yamaguchi potential specified by a scattering length $a_s$ and an effective
range ${r_0}_s$ (both in fm) in the $^1S_0$ channel, and $\Lambda$ separation
energy $B_\Lambda(^3_\Lambda n)$ (in MeV) obtained by solving the $\Lambda nn$
Faddeev equations with a separable Yamaguchi $\Lambda N$ spin-independent
interaction specified by a scattering length $a= -1.804$~fm and an effective
range $r_0=$ 2.5~fm.}
{\begin{tabular}{@{}p{1.cm}cp{1.cm}cp{1.cm}cp{1.cm}cp{1.cm}@{}} 
\toprule
 &  $a_s$ && ${r_0}_s$ && $B(^2n)$ && $B_\Lambda(^3_\Lambda n)$ &   \\
\colrule
 & \hphantom{0}5.4\hphantom{00}       && 1.75\hphantom{0}   && 2.23  && 0.39\hphantom{0}  & \\
 & \hphantom{0}5.4\hphantom{00}       && 2.25\hphantom{0}   && 2.79  && 0.27\hphantom{0}  & \\
 & \hphantom{0}5.4\hphantom{00}       && 2.881  && 4.98  && 0.16\hphantom{0}  & \\
 & \hphantom{0}6.0\hphantom{00}       && 2.881  && 2.86  && 0.11\hphantom{0}  & \\
 & \hphantom{0}7.0\hphantom{00}       && 2.881  && 1.64  && 0.06\hphantom{0}  & \\
 & \hphantom{0}9.0\hphantom{00}       && 2.881  && 0.80  && 0.01\hphantom{0}  & \\
 & \hphantom{0}13.0\hphantom{00}      && 2.881  && 0.32  && 0.003 & \\
 & \hphantom{0}17.612                 && 2.881  && 0.16  && $-$   & \\
 & $-$17.612                          && 2.881  && $-$   && $-$  & \\
\botrule
\end{tabular}}
\end{table}
 
The comparison between $^3_\Lambda n$ and the excitation energy of
$^4_\Lambda$H, $1^+_{exc} - 0^+_{g.s.} \approx 1.1$\,MeV, has been done using
the fact that the $\Lambda N - \Sigma N$ transition 
is dominated by the $G$ matrix effective interaction
devised by Akaishi {\it et al.}~\cite{Aka00} from the Nijmegen soft-core
interaction model NCS97~\cite{Rij99},
\begin{equation}
V_{\Lambda\Sigma}=(\bar V_{\Lambda\Sigma}+\Delta_{\Lambda\Sigma} \,
\vec S_N\cdot\vec S_Y)\sqrt{4/3}\,\, \vec t_N\cdot\vec t_{\Lambda\Sigma} \, ,
\end{equation} 
where $\vec t_{\Lambda\Sigma}$ converts a $\Lambda$ to $\Sigma$ in
isospin space and 
$\bar V_{\Lambda\Sigma}$ and $\Delta_{\Lambda\Sigma}$ are derived from
the Nijmegen model. The $1^+_{exc} - 0^+_{g.s.}$ excitation energy cannot
be reconciled with theory without substantial $\Lambda N - \Sigma N$ 
contribution~\cite{Aka00}. Such contribution is also relevant in neutron-rich
hypernuclei~\cite{Gal13}.

Focusing on the $(I,J^P)=(1,1/2^+)$ $^3_\Lambda$H state, particularly
relative to the $(0,1/2^+)$ $^3_\Lambda$H ground state, it was used the SU(4) limit of
nuclear core dynamics, in which the dineutron becomes bound and degenerate
with the deuteron, and where the difference in $\Lambda$ separation
energies of $(1,1/2^+)$ $^3_\Lambda$H and $(0,1/2^+)$ $^3_\Lambda$H 
is given by $\delta B_\Lambda=$ 0.26~MeV. Charge independence arguments allow
to estimate the $\Lambda$ separation energy in this hypothetical bound $^3_\Lambda n$
with respect to the bound dineutron core to be 0.39$\pm$0.05~MeV. Next, by
solving the $\Lambda nn$ Faddeev equations a $\Lambda n$
spin-independent Yamaguchi separable interaction was fitted. It reproduces 
$B_\Lambda(^3_\Lambda n)=$ 0.39~MeV, with $B(^2n)=$ 2.23~MeV as in the
deuteron. For the $nn$ interaction a Yamaguchi separable potential 
determined by the isoscalar $NN$ low-energy parameters, $a_s=$ 5.4~fm and 
${r_0}_s=$ 1.75~fm, was used. It gives rise to a dineutron binding energy $B(^2n)=$ 2.23~MeV, 
which equals the deuteron binding energy in the SU(4) limit.
Finally, a series of $\Lambda nn$ Faddeev calculations were performed by
keeping the $\Lambda n$ interaction fixed, but breaking SU(4) progressively
by varying the $nn$ interaction to reach $a_s=-17.612$~fm and ${r_0}_s=$ 2.881~fm,
as appropriate in the real world to the unbound dineutron. This is documented
in Table~4. 

Table~4 demonstrates the behavior of the dineutron binding energy
$B(^2n)$ and the $^3_\Lambda n$ binding energy 
$B(^3_\Lambda n)=B(^2n)+B_\Lambda(^3_\Lambda n)$ upon varying the $NN$
low-energy scattering parameters from the values given by the isoscalar $pn$
interaction down to the empirical values for the isovector $nn$ interaction.
This is done in two stages. First, increasing the effective range while
keeping the scattering length fixed, $B(^2n)$ increases whereas 
$B_\Lambda(^3_\Lambda n)$ steadily decreases. In the second stage,
keeping the effective range fixed at its final empirical $nn$ value, the
scattering length is varied by increasing it and then crossing from a
large positive value associated with a loosely bound dineutron to the
empirical large negative value of $a_{nn}$ associated with a virtual
dineutron. It can be seen how the dineutron binding energy, $B(^2n)$, 
decreases steadily when increasing the $NN$ scattering length for a fixed value
of the effective range. It is also observed how the binding increases for a fixed scattering length
when increasing the effective range as long as $a > r_0$. During this
stage $B_\Lambda(^3_\Lambda n)$ also decreases until
$^3_\Lambda n$ is no longer bound. This behavior makes evident the
anti-Borromean character of the $^3_\Lambda n$ system. Note that for
a Coulomb interaction, the $\Lambda - (nn)$ effective interaction would 
be roughly independent of the radius of $nn$ thanks to the Gauss theorem.
Thus, for example, for an attractive $\Lambda N$ potential of exponential shape the folding
on a spherical shell or on a sphere looks more favorable than concentrating
all strengths at the center, precisely the opposite behavior to that observed
in the last two columns of Table~4.

It is interesting to try to understand the behavior of the binding energy of the
two-body system shown in Table~4 when one varies  the low-energy
parameters $a$ and $r_0$~\cite{Pre62}. The two-body amplitude
for positive energies is given by 
\begin{equation}
t(E)=e^{i\delta}sin\delta/k=1/(k\, cotg\delta-ik) \, ,  
\label{eq91} 
\end{equation}
where the energy is $E=k^2/m$ with $m$ the mass of the neutron and
the effective-range expansion is
\begin{equation}
k \, cotg\delta=-1/a+k^2r_0/2+...   \, .  
\label{eq92}
\end{equation}
The bound states of the system are the poles of
the two-body amplitude, Eq.~(\ref{eq91}), when $k\to i\kappa$ so that
the energy of the bound state is $E=-\kappa^2/m$. If
one uses the effective-range expansion keeping only
the first two terms shown in Eq.~(\ref{eq92}) the 
position of the bound state is determined by
\begin{equation}
-1/a-\kappa^2r_0/2+\kappa=0 \, ,       
\label{eq93}
\end{equation}
which leads to
\begin{equation}
\kappa=\frac{1-\sqrt{1-2r_0/a}}{r_0} \, .
\label{eq94}
\end{equation}

If $2r_0/a<<1$ then $\kappa\to 2r_0/a^2$ so
that if $a$ is kept constant the binding energy increases
when $r_0$ increases. If one uses in Eq.~(\ref{eq94}) the values of 
$a$ and $r_0$ given in the first and second lines of Table~4 
one gets respectively $E= 2.24$ and 2.87 MeV quite close to 
the exact values. Using the values of the third line, Eq.~(\ref{eq94})
breaks down so that one needs to include the higher order terms
in the effective-range expansion~(\ref{eq92}) which are automatically 
included if one uses the separable potential and leads to the result
of the third line of Table~4. 

If one now keeps $r_0$ constant and increase $a$ as
in lines 4, 5, etc, then one sees from Eq.~(\ref{eq94}) that 
when $a\to\infty$ then
$\kappa\to 0$ and consequently $B\to 0$  as seen
in Table~4.

It is worth noting in Table~4 that
the dissociation of $^3_\Lambda n$ occurs while the dineutron is still
bound, although quite weakly. The final result of no $^3_\Lambda n$ bound 
state, for a virtual dineutron and $\Lambda N$ low-energy scattering parameters 
listed in the caption to Table~4, should come at no surprise given that a 
considerably larger-size $\Lambda N$ scattering length was found to be required 
in the Faddeev calculations to bind $^3_\Lambda n$, specifically $a=-4.492$~fm 
and $a=-5.930$~fm. Although a particular value of 2.5 ~fm for the $\Lambda N$ 
effective range was used in this demonstration, similar results are obtained 
for other reasonable choices of the $\Lambda N$ effective range~\cite{Gal14}.

\section{The $\Lambda\Lambda nn$ ($^4_{\Lambda\Lambda} n$) system}
\label{s3}

We discuss now different theoretical calculations where the $^4_{\Lambda\Lambda} n$ system
has been studied.

\subsection{A possible $\Lambda\Lambda nn$ bound state}
\label{ss31}

In Ref.~\refcite{Ric15} the possible existence of a $^4_{\Lambda\Lambda} n$ 
bound state was studied using a simple Gaussian variational method with
attractive interactions for the $nn$, $n\Lambda$ and
$\Lambda\Lambda$ subsystems. The chosen form of the interaction was
either a single Yukawa attractive term
\begin{equation}
V(r)=-g\, e^{-\mu r},
\end{equation} 
or a Morse parametrization
\begin{equation}
V(r)=g\, [e^{-2\mu(r-R)}-2e^{-\mu(r-R)}]\, ,
\end{equation} 
with $R=0.6$~fm. The two parameters $g$ and $\mu$ were adjusted to the two low-energy
parameters $a$ and $r_0$ of the various two-body interactions. In one case, these 
parameters were chosen from the Nijmegen-RIKEN ESC08 potential~\cite{Rij10,Rij13} 
and in another case from chiral effective field theory~\cite{Pol07,Hai13}. The
parameters are discussed in Ref.~\refcite{Ric15}.

It was found that the $^4_{\Lambda\Lambda} n$ system misses binding by a very small
amount with the Nijmegen-RIKEN parameters, but becomes bound by about
1~MeV with the chiral effective field theory parameters.

\subsection{The effect of repulsion in the $\Lambda\Lambda nn$ state}
\label{ss32}
\begin{table}[t]
\tbl{$S$ wave two-body channels contributing to the $(I,J^P)=(1,0^+)$ $\Lambda\Lambda nn$ system.}
{\begin{tabular}{@{}cccp{0.05cm}|p{0.05cm}ccc@{}}
  \toprule
	$V_{12}$                      & $-$ & $V_{34}$                       & & &
	$V_{13}$                      & $-$ & $V_{24}$\\ \colrule
	$nn$ $(i,j)=(1,0)$            & $-$ & $\Lambda\Lambda$ $(i,j)=(0,0)$ & & &
	$n\Lambda$ $(i,j)=(1/2,0)$    & $-$ & $n\Lambda$ $(i,j)=(1/2,0)$ \\	
	                              &     &                                &  & &
	$n\Lambda$ $(i,j)=(1/2,1)$    & $-$ & $n\Lambda$ $(i,j)=(1/2,1)$ \\	
	\botrule
 \end{tabular}}
 \end{table}
\begin{table}[b]
\tbl{Low-energy parameters and parameters of the local central Yukawa-type potentials 
given by Eq.~\eqref{eq21} for the $NN$, $\Lambda N$, and $\Lambda \Lambda$ spin-isospin $(i,j)$ 
two-body channels contributing to the $(I,J^P)=(1,0^+)$ $\Lambda\Lambda nn$ state. $A$ and $B$ are in MeV fm,
$\mu_A$ and $\mu_B$ are in fm$^{-1}$, and $a$ and $r_0$ are in fm.}  
{\begin{tabular}{@{}cccccccccc@{}}
\toprule 
& Ref. & \hphantom{00}$(i,j)$ & $A$ & $\mu_A$ & $B$ & $\mu_B$  & $a$ & $r_0$ &\\
\colrule
\multirow{1}{*}{$NN$} & ~\refcite{Gib90} & \hphantom{00}$(1,0)$  &  $513.968$  & $1.55$  & $1438.72$ & $3.11$ & $-23.56$ & $2.88$ &\\
\multirow{2}{*}{$\Lambda N$}& \multirow{2}{*}{~\refcite{Nag19}} & $(1/2,0)$ &   $416$\hphantom{000}  & $1.77$  & $1098$\hphantom{000} & $3.33$ & \hphantom{0}$-2.62$  & $3.17$  &\\ 
&& $(1/2,1)$ &   $339$\hphantom{000}  & $1.87$  & $968$\hphantom{000} & $3.73$ & \hphantom{0}$-1.72$  & $3.50$  &\\ \hline
\multirow{2}{*}{$\Lambda \Lambda$} & ~\refcite{Nag15} & \hphantom{00}$(0,0)$ &   $121$\hphantom{000}  & $1.74$  & $926$\hphantom{000} & $6.04$ & \hphantom{0}$-0.85$& $5.13$ &\\ 
 & ~\refcite{Sas18} & \hphantom{00}$(0,0)$ &   $207.44$\hphantom{0}  & $1.87$  & $627.6$\hphantom{0} & $3.63$ & \hphantom{0}$-0.62$& $7.32$ &\\ \botrule
\end{tabular}}
\end{table}

In Ref.~\refcite{Gar17} the possible existence of a $^4_{\Lambda\Lambda} n$ bound state 
was studied using a generalized Gaussian variational method~\cite{Vij09,Via09} 
with interactions that contain both attraction and repulsion 
for the $nn$, $n\Lambda$ and $\Lambda\Lambda$ subsystems. 
We summed up in Table~5 the different two-body channels contributing 
to the $(I,J^P)=(1,0^+)$ $\Lambda\Lambda nn$ state. The chosen form 
of the interaction was local central Yukawa-type Malfliet-Tjon 
potentials~\cite{Mal70},
\begin{equation}
V(r)=-A\frac{e^{-\mu_A r}}{r}+B\frac{e^{-\mu_B r}}{r},
\label{eq21}
\end{equation} 
where the four free parameters $A$, $B$, $\mu_A$, and $\mu_B$ were 
determined by fitting the low-energy data and the phase-shifts of each 
channel as given in the update of the strangeness 
$-1$~\cite{Nag19} and $-2$~\cite{Nag15,Rij16} ESC08c Nijmegen 
potentials. The low-energy data and the parameters of these models, together
with those of the $NN$ interaction of Ref.~\refcite{Gib90}, are summed up in Table~6.

The improved description of the observables of the two- and three-body subsystems 
as compared to Ref.~\refcite{Ric15}, with special reference to 
the introduction of the repulsive barrier for the $^1S_0$ $NN$ partial wave
relevant for the study of the tritium binding energy (see Table II 
of Ref.~\refcite{Mal70}), leads to a $\Lambda\Lambda nn$ four-body 
state above threshold. Besides, it can not get bound by reliable modifications of the
two-body subsystem interactions. 

In order to see how far the $^4_{\Lambda\Lambda} n$ system is from being bound, 
the dependence of the binding energy on the strength of
the attractive part of the different two-body interactions entering 
the four-body problem has been studied. For this purpose, the attractive 
part of the Malfliet-Tjon potential was multiplied by a fudge factor $g_{B_1B_2}$ as,
\begin{equation}
V^{B_1B_2}(r)=-g_{B_1B_2}A\frac{e^{-\mu_A r}}{r}+B\frac{e^{-\mu_B r}}{r} \, .
\label{Eqpot}
\end{equation} 
The system hardly gets bound for a reasonable increase of the strength of 
the $\Lambda\Lambda$, $g_{\Lambda\Lambda}$, interaction. 
Although one cannot exclude that the genuine $\Lambda\Lambda$ interaction in 
dilute states as the one studied here could be slightly stronger that the one 
reported in Ref.~\refcite{Nag15}, however, one needs $g_{\Lambda\Lambda} \ge 1.8$ to
get a four-body bound state, which destroys the agreement with the ESC08c Nijmegen 
$\Lambda\Lambda$ phase shifts. Note also that this is a very sensitive 
parameter for the study of double-$\Lambda$ hypernuclei~\cite{Nem03}
and this modification would produce an almost $\Lambda\Lambda$ bound state
in free space, in particular it would give rise to $a_{^1S_0}^{\Lambda\Lambda} = -29.15$~fm
and ${r_0}_{^1S_0}^{\Lambda\Lambda} = 1.90$~fm.

The four-body system becomes bound taking a multiplicative factor $1.2$ in the $NN$ interaction.
However, such modification would make the $^1S_0$ $NN$ potential as strong as the effective 
central $^3S_1$ interaction of Ref.~\refcite{Mal69} reproducing the deuteron binding energy and 
thus the singlet $S$ wave would develop a dineutron bound state, $a_{^1S_0}^{NN} = 6.07$~fm
and ${r_0}_{^1S_0}^{NN} = 1.96$~fm. A similar situation was encountered 
in Sect.~\ref{ss23} when a $\Lambda nn$ bound state was generated.

The situation is slightly different when dealing with the $\Lambda N$ interaction, which is dominant
because it contributes four times in the four-body problem, in contrast to the $nn$ or $\Lambda\Lambda$ 
interactions that only contribute one time. In this case, one uses
a multiplicative common factor $g_{N\Lambda}$ for the attractive part of the two $\Lambda N$
partial waves, $^1S_0$ and $^3S_1$. The four-body system develops a bound state for $g_{N\Lambda}=1.1$,
giving rise to $\Lambda N$ low-energy parameters: $a_{^1S_0}^{\Lambda N} = -5.60$~fm,
${r_0}_{^1S_0}^{\Lambda N} = 2.88$~fm, $a_{^3S_1}^{\Lambda N} = -2.91$~fm, and
${r_0}_{^3S_1}^{\Lambda N} = 2.99$~fm. They are far from the values constrained by the
existing experimental data. In particular, these scattering lengths point to the
unbound nature of the $\Lambda\Lambda nn$ system based on the hyperon-nucleon
interactions derived from chiral effective field theory in Ref.~\refcite{Hai19}, because it is
less attractive: $a_{^1S_0}^{\Lambda p} \in [-2.90, -2.91]$ fm and  
$a_{^3S_1}^{\Lambda p} \in [-1.40,-1.61]$ fm (see Table 1 of Ref.~\refcite{Hai19}). 
Similar results were obtained in Sect.~\ref{ss23} when
trying to get a $\Lambda nn$ bound state, destroying the agreement with the
known experimental data of the two-body subsystems.

The results are not very sensitive to the strength of the $\Lambda\Lambda$ interaction~\cite{Gar19}.
The calculation can be repeated using the latest $\Lambda\Lambda$ interaction derived by the 
lattice HAL QCD Collaboration~\cite{Sas18}. The parameters of the $\Lambda\Lambda$ HAL QCD potential
are given in the last row of Table~6. Although the $\Lambda\Lambda$ interaction
of Ref.~\refcite{Sas18} is slightly more attractive than that of the 
Nijmegen ESC08c potential~\cite{Nag15,Rij16}, the $\Lambda \Lambda nn$ state remains unbound.
The more attractive character of the HAL QCD $\Lambda\Lambda$ interaction can be easily tested
by trying to generate a $\Lambda\Lambda nn$ bound state with a multiplicative factor
for the $\Lambda\Lambda$ interaction in the attractive term of Eq.~\eqref{Eqpot}.
While with the model of Ref.~\refcite{Nag15} it is necessary a multiplicative factor 
$g_{\Lambda\Lambda}=1.8$ to get a bound state, with that of Ref.~\refcite{Sas18} the bound 
state is developed for $g_{\Lambda\Lambda}=1.6$.

Thus, the $^4_{\Lambda\Lambda} n$ does not seem to be Borromean, a four-body bound
state without two- or three-body stable subsystems. 
As clearly explained in Ref.~\refcite{Ric15}, the window of
Borromean binding is more an more reduced for potentials 
with harder inner cores. It is worth to note that Ref.~\refcite{Ric15} 
uses an intermediate version of the chiral effective interaction by the J\"ulich group~\cite{Pol07,Hai13}, 
presumably the latest at that time. However, the next iteration by the J\"ulich group led to drastic changes 
of the effective range for some of the baryon-baryon interactions~\cite{Hai16}.
This dichotomy is apparent also sometimes for the Nijmegen soft core potentials~\cite{Gas12},
in their quest for refinements, they combined scattering data and information from hypernuclei 
in which some medium corrections are perhaps at work. On the other hand, the very weakly bound 
systems are very dilute and may not experience medium corrections.
This would have to be considered in future studies of hypernuclei.

Finally, it is worth to note that an unbound result for the 
$^4_{\Lambda\Lambda} n$ has also been
reported in Ref.~\refcite{Lek14}. In this case the authors made use
of repulsive Gaussian-type potentials for any of the
two-body subsystems (see the figure on page 475) what does not allow
for the existence of any bound state.

\subsection{Pionless effective field theory and the 
$^3_{\Lambda\Lambda} n$ and $^4_{\Lambda\Lambda} n$ systems}
\label{ss33new}

Recently, Contessi {\it et al.}~\cite{Con19} used a stochastic variational method to
perform the first comprehensive pionless effective field theory study of $\Lambda\Lambda$ hypernuclei
with $A\le 6$. In addition to the interaction terms involved in the
description of single--$\Lambda$ hypernuclei,
a two-body $\Lambda\Lambda$ contact term
constrained to the $\Lambda\Lambda$ scattering length $a_{\Lambda\Lambda}$ was considered. 
A range of values compatible with $\Lambda\Lambda$ correlations
observed in relativistic heavy ion collisions as well as a three-body
$\Lambda\Lambda N$ contact term constrained to the binding energy of 
$^6_{\Lambda\Lambda}$He, the Nagara event~\cite{Tak01}, were used. It was 
found that the neutral three-body and four-body systems $^3_{\Lambda\Lambda}n$
and $^4_{\Lambda\Lambda}n$ are unbound by a large margin.

\subsection{A $\Lambda\Lambda (nn) - \Xi^- p (nn)$ three-body model}
\label{ss33}
In Ref.~\refcite{Gar19} the $\Lambda\Lambda nn - \Xi^- p nn$
coupled channel system was addressed by means of a three-body model.
The dineutron $(nn)$ was treated as an elementary 
particle with mass $m_{(nn)}=2m_n$, isospin 1, and spin 0 with
two-body interactions given by Yamaguchi separable potentials~\cite{Yam54}. 
Thus, a similar model to that proposed in Ref.~\refcite{Gar16} to search for 
resonances of the $\Lambda\Lambda N - \Xi NN$ system arises. If one of the
nucleons in the lower and upper channels is replaced by a dineutron, $N\to (nn)$, 
the equations of Ref.~\refcite{Gar16} are similar to those of this system. The differences
originate from the fact that in the $\Lambda\Lambda N - \Xi NN$ system 
two of the three particles in the upper channel are identical while in the
$\Lambda\Lambda (nn) - \Xi^- p (nn)$ system 
the three particles in the upper channel are different.

For all the uncoupled interactions one assumes separable potentials
of the form,
\begin{equation}
V_i^\rho = g_i^\rho\rangle \lambda_i^\rho\langle g_i^\rho \, .
\label{eq2}
\end{equation}
In the case of the $(i,j)=(0,0)$ two-body channel, responsible for the channel 
coupling $\Lambda\Lambda (nn) - \Xi^- p (nn)$, it was used a separable 
interaction of the form,
\begin{equation}
V_{1}^{\rho\sigma} = g_1^\rho\rangle \lambda_1^{\rho-\sigma}\langle g_1^\sigma \, .
\label{eq5}
\end{equation}
For the separable potentials of Eqs.~\eqref{eq2} and~\eqref{eq5}
one uses Yamaguchi form factors, i.e.,
\begin{equation}
g(p)=\frac{1}{\alpha^2+p^2} \, ,
\label{eq11}
\end{equation}
and thus, for each two-body channel one has to fit the
two parameters $\alpha$ and $\lambda$.

The $\Xi^- p (nn) \to \Lambda\Lambda (nn)$ process 
occurs with quantum numbers $(I,J^P)=(1,0^+)$ so that,
restricting the calculation to $S$ waves,
the contributing two-body channels in the three-body model are: the
$(nn)p$ channel $(i,j)=(1/2,1/2)$, the 
$(nn)\Lambda$ channel  $(i,j)=(1,1/2)$, the 
$(nn)\Xi^-$ channel $(i,j)=(3/2,1/2)$, and the $\Lambda\Lambda - \Xi^- p$
channel $(i,j)=(0,0)$.

In the case of the $(nn)p$ subsystem with quantum numbers
$(i,j)=(1/2,1/2)$, the tritium channel, for a given value of the range
$\alpha$ the tritium binding
energy, $8{.}48$~MeV, determines the strength $\lambda$.
The value of $\alpha$ is determined from the binding energy of 
$^4{\rm He}$, 28.2~MeV, through the solution of the three-body system $(nn)pp$.
The parameters of this model are given in the first row of Table~7.
\begin{table}[t]
\tbl{Parameters of the different separable potential models
for the uncoupled partial waves: $\alpha$ (in fm$^{-1}$) and $\lambda$
(in fm$^{-2}$).} 
{\begin{tabular}{@{}p{1.cm}ccccccp{1.cm}@{}} 
\toprule
& Model & Subsystem  & $(i,j)$ &  $\alpha$ & $\lambda$  && \\
\colrule
&& $(nn)p$                                 &  (1/2,1/2)      & 1.07   & $-$0.5444 & \\
&\multirow{2}{*}{1} & $(nn)\Lambda$        &  \hphantom{00}(1,1/2)        & \multirow{2}{*}{1.0\hphantom{0}}    & $-$0.1655 && \\
&                   & \hphantom{0}$(nn)\Xi^-  $        &  (3/2,1/2)      &                         & $-$0.2904 && \\
&\multirow{2}{*}{2} & $(nn)\Lambda$        &  \hphantom{00}(1,1/2)        & \multirow{2}{*}{2.0\hphantom{0}}    & $-$1.1560 && \\
&                   & \hphantom{0}$(nn)\Xi^-  $        &  (3/2,1/2)      &                         & $-$1.7719 && \\
&\multirow{2}{*}{3} & $(nn)\Lambda$        &  \hphantom{00}(1,1/2)        & \multirow{2}{*}{3.0\hphantom{0}}    & $-$3.9450 && \\
&                   & \hphantom{0}$(nn)\Xi^-  $        &  (3/2,1/2)      &                         & $-$5.4162 && \\ \botrule
									\end{tabular}} 
\end{table}
\begin{table}[b]
\tbl{Parameters of the different separable potential models
for the coupled partial wave $(i,j)=(0,0)$: $\alpha_1^{\Lambda\Lambda}$, 
$\alpha_1^{\Xi N}$ (in fm$^{-1}$), $\lambda_1^{\Lambda\Lambda}$,
$\lambda_1^{\Xi N}$, and $\lambda_1^{\Lambda\Lambda - \Xi N}$
(in fm$^{-2}$).}
{\begin{tabular}{@{}ccccccccc@{}} 
\toprule
& Model  &    $\alpha_1^{\Lambda\Lambda}$ & $\lambda_1^{\Lambda\Lambda}$
& $\alpha_1^{\Xi N}$  & $\lambda_1^{\Xi N}$  & $\lambda_1^{\Lambda\Lambda - \Xi N}$
&     \\
\colrule
& A & 1.3465            & $-$0.1390 & 1.1460            & $-$0.3867             & 0.0977            & \\
& B & 1.25\hphantom{00} & $-$0.0959 & 4.287\hphantom{0} & 1.302\hphantom{0}     & 1.243\hphantom{0} & \\ \botrule
\end{tabular}}
\end{table}

In the case of the $(nn)\Lambda$ subsystem with quantum numbers
$(i,j)=(1,1/2)$, the two parameters of the interaction were fitted
to the ground state and spin-excitation energies of the
$^4_\Lambda{\rm H}$ hypernucleus. It is considered as a three-body
system $(nn)p\Lambda$ with quantum numbers $(I,J^P)=(1/2,0^+)$. For
the $(nn)p$ subsystem the interaction previously described was used, and
for the $p\Lambda$ the separable potentials for $j=0$ and $j=1$
constructed in Ref.~\refcite{Gar16}. Thus,
for a given value of the range $\alpha$,
the strength $\lambda$ is fitted to the binding energy 
of $^4_\Lambda{\rm H}$, 10.52~MeV~\cite{Gal16}.
In order to obtain the range $\alpha$ one calculates the 
binding energy of the excited state $(I,J^P)=(1/2,1^+)$, 9.43~MeV~\cite{Gal16}, obtaining
for $\alpha$ = 1, 2, and 3~fm$^{-1}$ the values 9.93, 9.81 and 9.77~MeV,
respectively, which are labeled as models 1, 2, and 3 in Table~7. As it is
well known, the $^4_\Lambda{\rm H}$ spin excitation is difficult to fit since it depends
strongly on the tensor force arising from the transition 
$\Lambda N - \Sigma N$~\cite{Hiy14,Gal14,Gal16}. Therefore,
larger values of $\alpha$ were not considered.

In the case of the $(nn)\Xi^-$ subsystem with quantum numbers
$(i,j)=(3/2,1/2)$, there is not any experimental information 
available to calibrate the separable potential model.
Recent calculations~\cite{Gac16,Fil17} have studied the $\Xi NN$ 
system based in the strangeness $-2$ Nijmegen ESC08c potential~\cite{Nag15,Rij16}.
They reported a bound state with a binding energy of 2.89~MeV.
Thus, this result has been used to obtain the strength $\lambda$ of the
separable potential taking the range $\alpha$ equal to that of the
$(nn)\Lambda$ subsystem. We give in Table~7 the parameters 
corresponding to the different models 1, 2, and 3.

In the case of the coupled $\Lambda\Lambda - \Xi^-p$ subsystem two different
approaches were used. Firstly, a recent lattice QCD study by the HAL QCD 
Collaboration~\cite{Sas18} with almost
physical quark masses ($m_\pi=146$~MeV and $m_K=525$~MeV). 
In this model the $H$ dibaryon was calculated through the coupled 
channel $\Lambda\Lambda - \Xi N$ system, appearing
as a sharp resonance just below the $\Xi N$ threshold~\cite{Sas18,Ino12}.
It was constructed a model giving similar $\Lambda\Lambda$ and $\Xi N$ phase shifts as
those of Ref.~\refcite{Sas18}. The 
parameters of this model are given in Table~8 as model A.
Secondly, it has been also 
considered the separable potential model of the $\Lambda\Lambda-\Xi N$ system
constructed in Ref.~\refcite{Gar16} which is based in the Nijmegen
ESC08c potential~\cite{Nag15,Rij16}. This model is given in Table~8 as model B. Of
course, in the $\Lambda\Lambda (nn) - \Xi^- p (nn)$ calculations one uses the parameters 
$\lambda_1^{\Lambda\Lambda-\Xi^-p}=\lambda_1^{\Lambda\Lambda-\Xi N}/\sqrt{2}$
and $\lambda_1^{\Xi^-p}=\lambda_1^{\Xi N}/2$.
\begin{table}[t]
\tbl{Energy eigenvalue of the $\Lambda\Lambda (nn) - \Xi^- p (nn)$
system (in MeV) measured with respect to the $\Xi^-pnn$ threshold.
The results in parenthesis are those of the uncoupled $\Xi^- t$ binding energy.}
{\begin{tabular}{@{}cccccc@{}}
\toprule 
& Model  & 1 &  2 & 3  & \\
\colrule
& A & $-12{.}80 - i\, 0{.}05$ ($-$12.73)  & $-13{.}46 - i\, 0{.}04$ ($-$13.37)    & $-13{.}52 - i\, 0.04$ ($-$13.43) & \\
& B & $-10{.}99 - i\, 0{.}06$ ($-$10.92)  & $-11{.}04 - i\, 0{.}07$ ($-$10.93)    & $-10{.}90 - i\, 0.07$ ($-$10.77) & \\
\botrule
\end{tabular}}
\end{table}
\begin{table}[b]
\tbl{$\Xi^-t$ scattering length (in fm).
The results in parenthesis are those of the uncoupled $\Xi^- t$ scattering length.}
{\begin{tabular}{@{}cccccc@{}}
\toprule 
& Model  & 1 &  2 & 3  & \\
\colrule
& A & $1{.}286 - i\, 0{.}005$ $(1{.}293)$  & $1{.}030 - i\, 0{.}003$ $(1{.}036)$    & $0{.}957 - i\, 0{.}003$ $(0{.}963)$ & \\
& B & $1{.}551 - i\, 0{.}015$ $(1{.}567)$  & $1{.}315 - i\, 0{.}016$ $(1{.}339)$    & $1{.}268 - i\, 0{.}018$ $(1{.}298)$ & \\
\botrule
\end{tabular}}
\end{table}

Table~9 shows the energy eigenvalue of the two models A$-$B
of the coupled $\Lambda\Lambda - \Xi N$ system 
and the three models 1$-$3 of the $(nn)\Lambda$ and $(nn)\Xi^-$ systems. 
In parentheses, it is shown the energy of the uncoupled $\Xi^- t$ system. As one
can see from this table the real part of the energy eigenvalue is
slightly below the energy of the uncoupled $\Xi^- t$ system and 
the imaginary part of the energy eigenvalue is roughly the 
difference between the uncoupled energy and the real part of 
the energy eigenvalue. Thus, this state appears as a narrow $\Xi^- t$
quasibound state decaying to $\Lambda\Lambda nn$. The reason 
for the narrow width of the $\Xi^- t$ state stems from the weakness of the
$\Lambda\Lambda - \Xi N$ transition potential~\cite{Nag15,Rij16,Sas18}, that on the other hand is also
responsible for the $H$ dibaryon appearing as a 
very sharp resonance just below the $\Xi N$ threshold~\cite{Ino12}.

Finally, we give in Table~10 the corresponding values of the
$\Xi^-t$ scattering lengths of the two models A$-$B
of the coupled $\Lambda\Lambda - \Xi N$ system 
and the three models 1$-$3 of the $(nn)\Lambda$ and $(nn)\Xi^-$ systems,
which may be of use in the calculation of the energy shift of the 
atomic levels of the $\Xi^-t$ atom.

\subsection{The uncoupled $\Xi^- p nn$ system}
\label{ss34}
The uncoupled $\Xi^- p nn$ system with quantum numbers
$(I,J^P)=(1,0^+)$ has been studied in Ref.~\refcite{Gar19} 
using a generalized Gaussian variational method~\cite{Vij09,Via09}
to look for a possible bound state. This system contains several 
bound states made of subsets of two- and three-body particles. 
It contains the deuteron, the tritium,
the $(i,j)=(1,1)$ $\Xi N$ bound state predicted by the Nijmegen potential~\cite{Nag15,Rij16}
with a binding energy of $1{.}56$~MeV, and the $(i,j)=(3/2,1/2)$ $\Xi NN$ bound state with a
binding energy of $2{.}89$~MeV reported in Refs.~\refcite{Gac16} and~\refcite{Fil17}. If 
there were a $\Xi^- p nn$ bound state, 
it would not be stable unless its binding energy exceeds $m_{\Xi^- p} - m_{\Lambda\Lambda} = 28{.}6$
MeV. Otherwise it would decay to $\Lambda\Lambda nn$. 
If its binding energy would be larger
than that of the tritium, it would appear as a $\Xi^- t$ resonance or quasibound state
decaying to $\Lambda\Lambda nn$.

\begin{table}[t]
\tbl{$S$ wave two-body channels contributing to the $(I,J^P)=(1,0^+)$ $\Xi^- p nn$ system.} 
{\begin{tabular}{@{}cccp{0.05cm}|p{0.05cm}ccc@{}}
  \toprule
	$V_{12}$                      & $-$ & $V_{34}$               & & &
	$V_{13}$                      & $-$ & $V_{24}$\\ \colrule
	$nn$ $(i,j)=(1,0)$            & $-$ & $p\Xi^-$ $(i,j)=(0,0)$ & & &
	$np$ $(i,j)=(1,0)$            & $-$ & $n\Xi^-$ $(i,j)=(1,0)$ \\	
	$nn$ $(i,j)=(1,0)$            & $-$ & $p\Xi^-$ $(i,j)=(1,0)$ & & &
	$np$ $(i,j)=(0,1)$            & $-$ & $n\Xi^-$ $(i,j)=(1,1)$\\
	\botrule
 \end{tabular}}
 \end{table}
\begin{table}[b]
\tbl{Low-energy parameters and parameters of the local central Yukawa-type potentials 
given by Eq.~\eqref{eq21} for the $\Xi N$ system contributing to the $(I,J^P)=(1,0^+)$ 
$\Xi^- p nn$ state. $A$ and $B$ are in MeV fm, $\mu_A$ and $\mu_B$ are in fm$^{-1}$, 
and $a$ and $r_0$ are in fm.}
{\begin{tabular}{@{}cccccccccc@{}} 
\toprule
& Ref. & $(i,j)$ & $A$ & $\mu_A$ 
& $B$ & $\mu_B$  & $a$ & $r_0$ & \\
\colrule
\multirow{4}{*}{$\Xi N$}
&~\refcite{Sas18} & \multirow{2}{*}{$(0,0)$}              &  $161.38$  & $1.17$  & $197.5$   & $2.18$ & $-$ & $-$ &\\
&~\refcite{Nag15} &                                       &  $120$\hphantom{00}     & $1.30$  & $510$\hphantom{0}     & $2.30$ & $-$ & $-$ &\\
&~\refcite{Nag15} & $(1,0)$                               &  $290$\hphantom{00}     & $3.05$  & $155$\hphantom{0}     & $1.60$ & $0.58$  & $-2.52$ &\\
&~\refcite{Nag15} & $(1,1)$                               &  $568$\hphantom{00}     & $4.56$  & $425$\hphantom{0}     & $6.73$ & $4.91$  & \hphantom{0}$0.53$  & \\
\botrule
\end{tabular}}
\end{table}

To perform this study one needs the $\Xi N$ in three different partial waves.
We show in Table~11 the different two-body channels 
contributing to the $(I,J^P)=(1,0^+)$ $\Xi^- p nn$ state.
Ref.~\refcite{Gar19} presents two different calculations.
Firstly, the full set of $\Xi N$ interactions of the Nijmegen 
group~\cite{Nag15,Rij16} have been used parametrized by Malfliet-Tjon 
potentials as in Eq.~\eqref{eq21}.
Besides, as mentioned above, the HAL QCD Collaboration~\cite{Sas18} has recently derived 
a potential for the $(i,j)=(0,0)$ $\Lambda\Lambda -\Xi N$ coupled channel system with almost
physical quark masses. Thus, secondly, the calculation has been performed with
the HAL QCD potential~\cite{Sas18} for the $(i,j)=(0,0)$ $\Xi N$ channel.
The low-energy data and the parameters of the different $\Xi N$ interactions
are shown in Table~12.

Ref.~\refcite{Gar19} reported a $\Xi^- p nn$ bound state of $14{.}43$~MeV with
the $(i,j)=(0,0)$ $\Xi N$ HAL QCD interaction and $10.78$~MeV with the Nijmegen potentials. 
In both cases, the $(I,J^P)=(1,0^+)$ $\Xi^- p nn$ state lies below the 
lowest two-body threshold, $\Xi^- t$. Such state would decay to the $\Lambda\Lambda nn$ channel 
with a very small width, as shown in Sect.~\ref{ss33} and Ref.~\refcite{Gar18}.
The results are in close agreement with those obtained with the separable potential
three-body model shown in Table~9.
In all models the binding is larger than that of the
tritium and a slightly deeper bound state is obtained when using the HAL QCD interactions for the two-body
coupled channel $(i,j)=(0,0)$. By including the 
Coulomb $\Xi^- p$ potential the binding energies are increased roughly by $0{.}75$ MeV with the 
HAL QCD interaction and $0{.}53$ MeV with the Nijmegen potentials, driving to final binding
energies of $15{.}18$ MeV and $11{.}31$ MeV, respectively. 

The $(I,J^P)=(1,0^+)$ $\Xi^- p nn$ state has also been found to be bound in Ref.~\refcite{Hiy19} by using the 
AV8 NN potential of Ref.~\refcite{Wir84} and the full set of the ESC08c strangeness
$-2$ $\Xi N$ interactions of Refs.~\refcite{Nag15} and~\refcite{Rij16}, see Table 1 and Fig. 3b
of Ref.~\refcite{Hiy19}. It was found to be unbound with the full
set of the HAL QCD strangeness $-2$ $\Xi N$ interactions, when such interactions were still not available
in the literature, except for the $(i,j)=(0,0)$ channel~\cite{Sas18}.

\section{The $\Lambda\Lambda n$ ($^3_{\Lambda\Lambda} n$) system}
\label{s4}

The $\Lambda\Lambda N -\Xi NN$ coupled channel system has been studied in 
Refs.~\refcite{Gar13},~\refcite{Gac14},~\refcite{Gar15}, and~\refcite{Gac13}
by means of the CQCM~\cite{Val05,Vac05,Vij05} two-body interactions. Later on, Ref.~\refcite{Gar16} 
used separable potentials tuned to the low-energy parameters of the Nijmegen ESC08 baryon-baryon
interactions~\cite{Nag15,Rij16,Nag19}. Recently, Ref.~\refcite{Gac20} has studied this system
using a separable model of the available HAL QCD Collaboration potentials for the $(i,j)=(0,0)$ 
$\Lambda\Lambda - \Xi N$ channel~\cite{Sas18} and also the full set of 
$\Lambda\Lambda$ and $\Xi N$ HAL QCD baryon-baryon interactions with near-physical 
quark masses released when this review was already finished~\cite{Sas19}.

\subsection{Faddeev equations with quark model-based interactions}
\label{ss41}
\begin{table}[b]
\tbl{Different models for the strangeness $-$2 two-body interactions
characterized by the $\Lambda\Lambda$ scattering 
length, $a_{\Lambda\Lambda}$, and the uncoupled $\Lambda\Lambda$ scattering length
$a_{\Lambda\Lambda}^{\rm{U}}$, in fm. The last column stands for the binding energy of the $H$ dibaryon,
in MeV.} 
{\begin{tabular}{@{}p{1.0cm}cp{0.1cm}cp{0.1cm}cp{0.1cm}cp{1.0cm}@{}}
\toprule
& Model && $a_{\Lambda\Lambda}$ && $a_{\Lambda\Lambda}^{U}$ && $B_H$ & \\
\colrule
& A && 2.54 && $-$3.29 && 6.93 & \\
& B && 2.98 && $-$2.42 && 4.92 & \\
& C && 3.74 && $-$1.79 && 3.01 & \\
& D && 6.47 && $-$1.19 && 0.94 & \\
& E && 8.24 && $-$1.07 && 0.56 & \\
\botrule
\end{tabular}}
\end{table}

The Faddeev equations for the bound state problem of the coupled $\Lambda\Lambda N -\Xi NN$
system were for the first time solved in Refs.~\refcite{Gar13} and~\refcite{Gac14}. They were
later on generalized to include the coupling to $\Lambda\Sigma N$ and $\Sigma\Sigma N$ channels
in Ref.~\refcite{Gar15}. In both cases the CQCM baryon-baryon interactions~\cite{Val05,Vac05,Vij05} were used. 
In particular, for the strangeness $-1$ two-body systems one uses 
the set of $\Lambda N$ potentials discussed in Sect.~\ref{ss22}, satisfying 
the constraints $-1.41 \ge  a_{1/2,1} \ge -1.58$~fm and $-2.37 \ge a_{1/2,0} \ge -2.48$~fm for the
spin-triplet and spin-singlet $\Lambda N$ scattering lengths, respectively.
For the strangeness $-2$ two-body systems a set of interactions were devised
under the assumption that the $H$ dibaryon~\cite{Jaf77} has the lower limit mass
determined by the E373 experiment at KEK~\cite{Tak01} from the observation of
a $^6_{\Lambda\Lambda}$He double hypernucleus, $B_{\Lambda\Lambda} = 7.13 \pm 0.87$, 
up to the $\Lambda\Lambda$ threshold.
They are summed up in Table~13 characterized by the $\Lambda\Lambda$ scattering 
length, $a_{\Lambda\Lambda}$. In the same table it is also shown 
the uncoupled $\Lambda\Lambda$ scattering length, 
$a_{\Lambda\Lambda}^U$, of interest for the study
of double-$\Lambda$ hypernuclei, that it is calculated by dropping the
coupling to the $N\Xi$ channel. 

\begin{table}[t]
\tbl{Binding energy of the $(I,J^P)=(\frac{1}{2},\frac{1}{2}^+)$ 
strangeness $-2$ three-body state (in MeV) 
measured with respect to the corresponding $NH$ threshold (considering the
binding energy of the $H$ dibaryon in each particular model shown in Table~13)
for several models of the $YN$ interaction specified in Table~3 
and for models A and E of the strangeness $-2$ two-body interactions of Table~13.
The results in parenthesis were obtained neglecting the $N\Lambda\Sigma$ and 
$N\Sigma\Sigma$ channels.}
{\begin{tabular}{@{}ccccccc@{}}
\toprule
&    & $a^{\Lambda N}_{1/2,1}=-1.41$ & $a^{\Lambda N}_{1/2,1}=-1.46$ 
& $a^{\Lambda N}_{1/2,1}=-1.52$ & $a^{\Lambda N}_{1/2,1}=-1.58$ & \\ \colrule
\multicolumn{7}{c}{Model A}\\ \colrule
& $a^{\Lambda N}_{1/2,0}=-2.33$ & 0.335 (0.474) & 0.369 (0.512) & 0.407 (0.553) 
& 0.450 (0.601) & \\
& $a^{\Lambda N}_{1/2,0}=-2.39$ & 0.342 (0.482) & 0.377 (0.521) & 0.415 (0.562) 
& 0.458 (0.610) & \\
& $a^{\Lambda N}_{1/2,0}=-2.48$ & 0.363 (0.504) & 0.400 (0.545) & 0.438 (0.587) 
& 0.483 (0.636) & \\
\toprule
\multicolumn{7}{c}{Model E}\\ \colrule
& $a^{\Lambda N}_{1/2,0}=-2.33$ & 0.144 (0.222) & 0.171 (0.253) & 0.200 (0.287) 
& 0.236 (0.327) & \\
& $a^{\Lambda N}_{1/2,0}=-2.39$ & 0.150 (0.229) & 0.177 (0.261) & 0.207 (0.295) 
& 0.243 (0.335) & \\
& $a^{\Lambda N}_{1/2,0}=-2.48$ & 0.166 (0.246) & 0.194 (0.279) & 0.225 (0.314) 
& 0.263 (0.356) & \\
\botrule
\end{tabular}} 
\end{table}

Table~14 shows the results obtained for the binding energy of the 
$(I,J^P)=(\frac{1}{2},\frac{1}{2}^+)$ strangeness $-2$ three-body system
for the different models of the strangeness $-2$ two-body interactions constructed 
in Ref.~\refcite{Gac14} and shown in Table~13.
The results of the full $\Lambda\Lambda N -\Xi NN - \Lambda\Sigma N - \Sigma\Sigma N$
coupled channel calculation are given and in parenthesis, for comparison, the 
corresponding results when one neglects the $\Lambda\Sigma N$ and 
$\Sigma\Sigma N$ channels.
As one can see from this table the binding energy varies between 0.335 and 0.483 MeV (0.474 and 0.636 MeV
if the $\Lambda\Sigma N$ and $\Sigma\Sigma N$ channels are neglected)
for model A and between 0.144 and 0.263 MeV (0.222 and 0.356 if the $\Lambda\Sigma N$ and 
$\Sigma\Sigma N$ channels are neglected) for model E of the strangeness $-2$ two-body 
interactions. In all cases the three-body bound state is present slightly 
below the corresponding $NH$ threshold. It is important to note that 
the bound state only appears when one takes into account the 
coupling between the $\Lambda\Lambda N$ and $\Xi NN$ components, 
i.e., when one includes the $(i,j)=(0,0)$ two-body
$t^{\Lambda\Lambda - \Xi N}$ amplitude, otherwise the $\Lambda \Lambda N$ system
alone is not bound~\cite{Gar07,Gac07}. Thus, as predicted in Ref.~\refcite{Tan65},
this result is compatible with the non-existence of a stable $^3_\Lambda$H with isospin one.

Ref.~\refcite{Gaa13}, using the uncoupled $\Lambda\Lambda$ scattering
length of Ref.~\refcite{Gar13}, 
compared with results of three-body calculations of the $\Lambda\Lambda\alpha$ 
system in which either unrealistic separable potentials had been used for the 
two-body subsystems~\cite{Car97, Afn03} or the coupling $\Lambda\Lambda-N\Xi$ had been 
included only in an effective manner~\cite{Fil02,Fil03}.
From this comparison Ref.~\refcite{Gaa13} discusses a possible overbinding of the model
of Ref.~\refcite{Gar13} for the ${}_{\Lambda\Lambda}^6$He hypernucleus.
These arguments were challenged in Ref.~\refcite{Gac13}, see Table 15, showing that
the $(I,J^P)=(\frac{1}{2},\frac{1}{2}^+)$ strangeness $-2$ three-body state
survives to small uncoupled $\Lambda\Lambda$ scattering 
lengths satisfying the limits obtained in Ref.~\refcite{Ohn17}, 
while describing equally well the available experimental data.
\begin{table}[b] 
\tbl{Binding energy of the $(I,J^P)=(\frac{1}{2},\frac{1}{2}^+)$ 
strangeness $-2$ three-body state, $B_{\hat S=-2}$,
measured with respect to the $NH$ threshold for different values of the 
uncoupled $\Lambda\Lambda$ scattering length $a_{\Lambda\Lambda}^{U}$.
Energies are in MeV and the scattering length in fm.}
{\begin{tabular}{@{}p{1.cm}cp{0.1cm}cp{0.1cm}cp{1.cm}@{}}
 \toprule
 &$a_{\Lambda\Lambda}^{U}$ && $B_H$  && $B_{\hat S=-2}$  &\\
 \colrule
 &$-$ 3.3  &&  6.928   &&   0.577 &\\
 &$-$ 2.3  &&  6.191   &&   0.640 &\\
 &$-$ 1.3  &&  4.962   &&   0.753 &\\
 &$-$ 0.5  &&  3.250   &&   0.927 &\\
\botrule
\end{tabular}}
\end{table}

\subsection{Faddeev equations with separable potentials}
\label{ss42}

Despite the large amount of experimental and theoretical efforts, the existence
of the $H$ dibaryon remains inconclusive, see Ref.~\refcite{Fra19} for a recent update.
Experimental evidence disfavors large binding energies~\cite{Gal16}, as predicted in
Ref.~\refcite{Jaf77}, and the high statistics study of $\Upsilon$ decays at 
Belle~\cite{Kim13} found no indication of an $H$ dibaryon with a mass near 
the $\Lambda\Lambda$ threshold. Thus, Ref.~\refcite{Gar16} used the developments of 
Refs.~\refcite{Gar13} and~\refcite{Gac14} using separable potentials tuned to the 
low-energy parameters of the Nijmegen ESC08 baryon-baryon
interactions~\cite{Nag15,Rij16,Nag19} that give no indication of either a
bound state or a resonance in the strangeness $-$2 $(0,0^+)$ 
$\Lambda\Lambda - \Xi N$ two-body channel~\footnote{$\Lambda\Sigma N$ and $\Sigma\Sigma N$ 
channels were not considered because of the small 
contribution found in Ref.~\refcite{Gar15}}.
The three-body bound state problem was extended into the continuum
region to look for possible resonances above the $\Lambda\Lambda N$ threshold~\cite{Pea84}.
A resonance was obtained 23.41~MeV above the $\Lambda\Lambda N$ mass, just 12~keV below the $\Xi d$ threshold.
Due to the negligible $\Lambda\Lambda - \Xi N$ coupling predicted by the Nijmegen
potential, as already discussed is Sect.~\ref{ss33},
this resonance has a very small width of $\Gamma = 0.09$ MeV, so that 
it is practically a bound state. 

Ref.~\refcite{Gac20} made use of the more attractive HAL QCD $\Lambda\Lambda - \Xi N$ 
interaction~\cite{Sas18} predicting that the $H$ dibaryon could be a $\Lambda\Lambda$ 
resonance just below the $\Xi N$ threshold. It is worth to notice that similar results have been
obtained in a low-energy effective field theory study of the
$H$ dibaryon in $\Lambda\Lambda$ scattering~\cite{Yam16}.
The results were obtained by taking the nucleon mass as the
average of the proton and neutron masses and the $\Xi$ mass as the average 
of $\Xi^0$ and $\Xi^-$ masses. Thus, the $\Xi N$ and $\Xi NN$ thresholds are
25.6~MeV above the $\Lambda\Lambda$ and $\Lambda\Lambda N$ thresholds,
respectively. However, this mass difference is 32 MeV for the HAL QCD
results~\cite{Sas18}, since they use for the baryon masses the values
obtained from their lattice QCD study. Thus, one should keep in mind that 
the energy scale of Ref.~\refcite{Sas18} corresponds to that of
Fig.~1 of Ref.~\refcite{Gac20} multiplied by 1.25.
 
It was found a pole lying at $E=17.6- \, i \, 0.24$~MeV so that
the three-body resonance lies 8~MeV below the $\Xi NN$ threshold and
has a width of 0.48~MeV. The most intriguing feature of this state,
a $\Lambda\Lambda N$ resonance as seen from the lower component or
a $\Xi NN$ quasibound state as seen from the upper component, is
its very small width. It has been explained in Ref.~\refcite{Gac17} by
using first order perturbation theory, taking the $\Xi NN$ component
as the dominant one and the $\Lambda\Lambda N$ channel as the perturbation.
This is because the small effect induced by the $\Lambda\Lambda - \Xi N$
interaction in the pole position compared to the result obtained by 
neglecting the $(0,0^+)$ $\Lambda\Lambda - \Xi N$ interaction, in which 
case the three-body state would appear as a bound state of the $\Xi NN$ 
system. Thus, it clearly indicates that the lower three-body
channel effectively acts as a perturbation.

Let us finally mention that these calculations have been repeated~\cite{Gac20} with 
the full set of $\Lambda\Lambda$ and $\Xi N$ HAL QCD baryon-baryon interactions 
with near-physical quark masses~\cite{Sas19} released when this review was already finished.
In this case the Argand diagram of the $\Xi d$ system between 0 and 10
MeV above the $\Xi d$ threshold shows the typical counterclockwise
behavior of a resonant amplitude. If one neglects the coupling to the lower
$\Lambda\Lambda N$ channel the counterclockwise
behavior disappears, which shows that the resonance is due to the coupling
to the lower channel.

\section{Conclusions}

We have reviewed the work by several theoretical groups as regards the existence
of stable neutral baryonic systems with strangeness.
We have seen
that it is not possible to accommodate a $\Lambda nn$ bound state 
from our knowledge about nuclear and hypernuclear interactions.
In the case of the $\Lambda\Lambda nn$ system the conditions to
reach binding are somewhat closer since with
purely attractive interactions fitted to the low-energy data 
binding can be achieved for some models.
However, when the effect of repulsion is included no bound state is found. 

With the available two-body interactions that are adjusted to describe what is known 
about the two- and three-baryon subsystems, neither a $\Lambda\Lambda nn$ 
bound state nor a resonance is obtained. However, a possible
$\Xi^- t$ quasibound state with quantum numbers $(I,J)=(1,0)$ 
above the $\Lambda\Lambda nn$ threshold might exist in nature.
The stability of the state is increased by considering the Coulomb potential.
The different approaches to the $\Lambda\Lambda - \Xi N$ interaction
drive to similar results, the weakness of the 
$\Lambda\Lambda - \Xi N$ transition potential explaining the narrow width of the
$\Xi^- t$ quasibound state. 

The possible existence of a three-body $(I,J^P)=(1/2,1/2^+)$
bound, quasibound state or resonance with strangeness $-2$, pointed out by quark model-based and
Nijmegen and HAL QCD inspired separable potentials, has also been reviewed.  

\section*{Acknowledgments} 
This work has been partially funded by COFAA-IPN (M\'exico) and 
by Ministerio de Econom\'\i a, Industria y Competitividad 
and EU FEDER under Contract No. FPA2016-77177.

\end{document}